\documentclass[aps,prd,twocolumn,superscriptaddress,10pt,noshowpacs]{revtex4}  
\usepackage[active]{srcltx}
\usepackage{graphicx}
\usepackage[utf8]{inputenc}
\usepackage{amsmath}
\usepackage{times,txfonts}
\usepackage{float}

\begin{document}

\title{Position-dependent mass effects on a bilayer graphene catenoid bridge}

\author{J. E. G. Silva}
\affiliation{Centro de Ci\^{e}ncias e Tecnologia\\
Universidade Federal do Cariri\\
57072-270, Juazeiro do Norte, Cear\'{a}, Brazil}
\email{euclides.silva@ufca.edu.br}

\author{J. Furtado}
\affiliation{Centro de Ci\^{e}ncias e Tecnologia\\
Universidade Federal do Cariri\\
57072-270, Juazeiro do Norte, Cear\'{a}, Brazil}
\email{job.furtado@ufca.edu.br}

\author{A. C. A. Ramos}
\affiliation{Centro de Ci\^{e}ncias e Tecnologia\\
Universidade Federal do Cariri\\
57072-270, Juazeiro do Norte, Cear\'{a}, Brazil}
\email{antonio.ramos@ufca.edu.br}

\date{\today}

\begin{abstract}
We study the electronic properties of a position-dependent effective mass electron on a bilayer graphene catenoid bridge. We propose a position-dependent mass (PDM) as a function of both gaussian and mean curvature. The hamiltonian exhibits parity and time-reversal steaming from the bridge symmetry. The effective potential contains the da Costa, centrifugal and PDM terms which are concentrated around the catenoid bridge. For zero angular momentum states, the PDM term provides a transition between a reflectionless to a double-well potential. As a result, the bound states undergo a transition from a single state around the bridge throat into two states each one located at rings around the bridge. Above some critical value of the PDM coupling constant, the degeneracy is restored due to double-well tunneling resonance.

\end{abstract}

\maketitle


\section{Introduction}
\label{introduction}

Two dimensional structures, as the graphene \cite{geim,novoselov,katsnelson}, nanotubes \cite{nanotube}  and the phosphorene \cite{phosphorene} open a new venue to study the electron properties at low dimensional physics. The geometry of the graphene layer plays a pivotal role on the electronic structure.
The curvature at the tip of a conical layer produces topological phase \cite{furtado}, whereas  helical strips induces chiral properties \cite{dandoloff1,atanasovhelicoid,atanasov}. Fluctuations of the geometry produces the so-called pseudomagnetic fields \cite{ribbons}, whose effects can be seen at ripples \cite{contijo} and corrugated layers \cite{corrugated} . 

The electron Hamiltonian on the surface can be obtained from the $3D$ Hamiltonain by considering a small surface width $\epsilon$, writing the Hamiltonian in the tangent and the normal coordinates and then, taking the limit $\epsilon\rightarrow 0$ \cite{hjensen}. Starting with the $3D$ Schr\"{o}dinger Hamiltonian and appying this thin-layer squeezing method, one obtains a geometric potential, known as the da Costa potential \cite{costa,costa1,matsutani,luiz}.  The geometric da Costa potential depends on the squared of the gaussian and the mean curvatures and yields to an attractive potential. This method can also be extended to include external fields \cite{ferrari}, spin in a Pauli equation \cite{wang} and the Dirac equation on surfaces \cite{BJ}.

The geometry of the graphene layer can be used to develop new electronic devices. In Ref. \cite{wormhole,picak} a bridge connecting a bilayer graphene was devised using a nanotube. In order to obtain a smooth bridge, the Ref.\cite{Dandoloff} proposed a catenoid surface to describe the bilayer and the bridge using only one surface. This can be achieved due to the catenoid curvature which is concentrated around the bridge and vanishes asymptotically \cite{spivak}. The catenoid is a minimal surface, which is known to provide a stable graphitic structure \cite{terrones}.

In Ref.\cite{euclides}, we explored the effects of the geometry and of external electric and magnetic fields upon the graphene catenoid bridge. The da Costa potential provides a reflectionless attractive potential, whereas the symmetry with respect to the $z$ axis yields to a centrifugal repulsive term \cite{euclides}. The geometric potential exhibits a parity and time reversal symmetry which holds under the action of the magnetic field but it is broken by the external electric field \cite{euclides}. The magnetic field produces a double-well potential which leads to bound states located at symmetric rings around the catendoid throat. The electric field creates a difference between the asymptotic values of the effective potential on the upper to the lower layer. This effect suggests that the catenoid bilayer bridge could be used as a diode \cite{euclides}.

Besides the geometric potential, it is expected that the surface curvature may also produce position-dependent mass effects. Indeed, the curvature breaks the homogeneity of the lattice which compose the layer, thus modifying the electron effective mass. The PDM can be produced by $p-n$ junctions driven by curvature\cite{pnbilayer}, as well as phonon interactions\cite{sinner}. Although the position-dependent mass Hamiltonians in flat surfaces are widely studied \cite{pdm1,pdm2,pdm3,pdm4}, only recently an extension of the da Costa method including position-dependent mass effects on curved surfaces was proposed \cite{moraes}.  The inclusion of an effective mass of form $m^{*}\propto d^{-\alpha}$, where $d$ is the nanotube diameter, on a corrugated nanotube lead to significant modifications of the transport properties \cite{moraes}. In a cylinder, the gaussian curvature vanishes whereas the mean curvature is a constant $H=d^{-1}$. Thus, the position-dependent mass considered in Ref.\cite{moraes} is proportional to the a power of the corrugated mean curvature. 

In this work we study the effects of the position-dependent mass upon the electron on a catenoid bridge. In section \ref{section2}, we propose an isotropic position-dependent effective mass as a function of the gaussian and the mean curvatures. Then, we analyze qualitatively the features of the effective potential, such as its behaviour with respect to the parity, time-reversal symmetry and hermiticity. In section \ref{section3}, we obtain the bound states and the corresponding energy spectrum. Final remarks and perspectives are outlined in section \ref{remarks}.

\section{Position-dependent mass electron on a catenoid surface}
\label{section2}

\begin{figure}
\begin{center}
\label{catenoidsurface}
\includegraphics[scale=1.1]{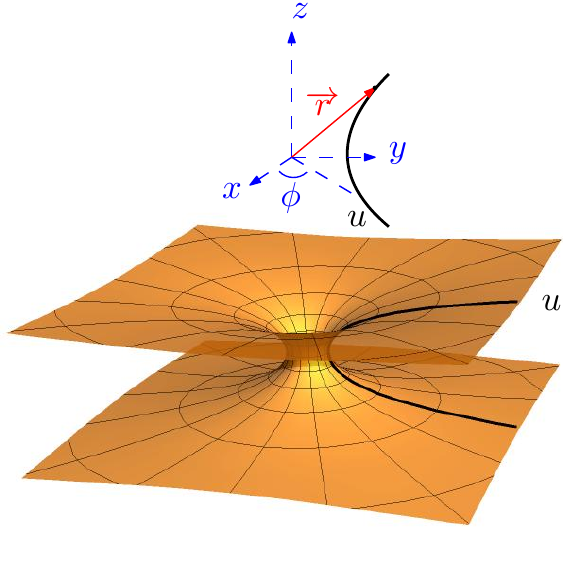}
\caption{Catenoid surface and coordinate system}
\end{center}
\end{figure}

In this section we introduce the geometry and the dynamics of the electron on the doublelayer catenoid bridge considering the effects of a curvature-dependent mass. As shown in fig.\ref{catenoidsurface}, the double layer graphene bridge is realized as a smooth minimal surface (least area) joining the two planes. Near the bridge throat, depicted in fig.\ref{catenoidsurface}, the symmetry about the $z$ axis and the minimal radius $R$ are shown.

After squeezing the electron wave-function on a surface, the spinless stationary Schr\"{o}dinger equation has the form
\begin{equation}
\label{constantmassschrodinger}
    -\frac{\hbar^2}{2m^*}\nabla^2 \Psi +V_{dc}\Psi=E\Psi,
\end{equation}
where $\nabla^{2}\Psi=\frac{1}{\sqrt{g}}\partial_{a}(\sqrt{g}g^{ab}\partial_{b}\Psi)$ is the Laplacian operator on the surface, $g^{ab}$ is the induced metric of the surface and $V_{dC} = - \frac{\hbar^2}{2m^{*}}(H^2-K)$ is a potential induced by the surface curvature, known as the da Costa potential \cite{costa}. The geometric potential depends both on the  the mean curvature $H$ and the Gaussian curvature $K$ \cite{costa}. 

Recently, the electronic properties of the electron on the bilayer catenoid bridge was investigated assuming a constant effective mass $m^*$ \cite{euclides}. Since the effective mass is also modified by the break of the lattice homogeneity driven by the curvature, we consider the modified position-dependent mass Schr\"{o}dinger equation in the form \cite{moraes}
\begin{eqnarray}
\label{pdmschrodinger}
(\mathcal{K}-V_{dC})\Psi=E\Psi,
\end{eqnarray}
where the position-dependent mass kinetic operator $\mathcal{K}$ is defined as
\begin{equation}
\label{modifiedoperator}
\mathcal{K}\Psi=-\frac{\hbar^2}{2}\left[\frac{1}{m^*}\nabla^{2}\Psi+\frac{1}{3}\nabla^{2}\left(\frac{1}{m^*}\right)\Psi+\partial^{j}\left(\frac{1}{m^*}\right)(\partial_i\Psi)\right],
\end{equation}
and $m^{*}=m^{*}(x)$. Note that for a constant effective mass $m^{*}$ the Eq.\eqref{constantmassschrodinger} is obtained.

Adopting a coordinate system formed from the the meridian $u=u(z)=R\sinh\left(z/R\right)$ and the parallel $\phi$, where $u\in (-\infty, \infty)$ and the parallel $\phi\in [0,2\pi)$, a point on the catenoid surface can be written as
\begin{equation}
\label{cilindricalcoordinates}
\vec{r}=\sqrt{R^2 + u^2}(\cos\phi\hat{i} +\sin\phi \hat{j})+ R\sinh^{-1}(u/R)\hat{k},
\end{equation}
where $R$ the radius of the catenoid bridge, as shown in figure \ref{catenoidsurface}.
In this coordinate system, the induced metric on the catenoid is $g_{uu}=1$ and $g_{\phi\phi}=R^2 +u^2$, and the da Costa potential 
\begin{equation}
    V_{dC} =-\frac{\hbar^2}{2m^{*}}\frac{R^2}{(R^2 + u^2)^2}. 
\end{equation}
It is worthwhile to mention that the da Costa potential $V_{dc}$ exhibits a parity-symmetrical potential well with respect to $u=0$. In addition, the da Costa potential vanishes asymptotically, reflecting the asymptotic flat geometry of the catenoid. Note that the curvature leads to an attractive potential which tends to trap the electron around the origin.     
 
The axial symmetry leads to the periodic behavior of the wave function in the form
\begin{equation}
\label{axialsymmetry}
\Psi(u,\phi)=\Phi(u)e^{i \nu\phi},
\end{equation}
where $\nu$ is the orbital quantum number. Substituting the Eq.\eqref{axialsymmetry} into Eq.\eqref{pdmschrodinger} yields to
\begin{eqnarray}
\label{meridianschrodingerequation}
\nonumber&-&\frac{\hbar^2}{2m^*}\left[\partial_u^2\Phi+\frac{u}{R^2+u^2}\partial_u\Phi-\frac{\nu^2}{R^2+u^2}\Phi\right]\\
\nonumber&&-\frac{\hbar^2}{6}\left[\partial_u^2\left(\frac{1}{m^*}\right)+\frac{u}{R^2+u^2}\partial_{u}\left(\frac{1}{m^*}\right)+\frac{1}{R^2+u^2}\partial_{\phi}^2\left(\frac{1}{m^*}\right)\right]\Phi\\
\nonumber&&-\frac{\hbar^2}{2}\left[\partial_u\left(\frac{1}{m^*}\right)(\partial_u\Phi)+\frac{i\nu}{R^2+u^2}\partial_{\phi}\left(\frac{1}{m^*}\Phi\right)\right]\\
&&-\frac{\hbar^2}{2m^*}\frac{R^2}{(R^2+u^2)^2}\Phi=E\Phi.
\end{eqnarray}

The Eq.\eqref{meridianschrodingerequation} depends on the position-dependence of the effective mass function $m^{*}$. In general, the effective mass may be anisotropic, as a result of a non-symmetric energy band \cite{moraes,anisotropicmass}. In Ref.\cite{moraes}, the authors explored the electronic properties of position-dependent Schr\"{o}dinger equation \eqref{pdmschrodinger} on a nanotube. For that purpose, they considered $m^{*}\propto d^{-\alpha}$, where $d$ is the nanotube diameter \cite{moraes}.

Here we propose that 
the curvature breaks the homogeneity of the system, and the consequence of such breaking leads to a modification in the effective mass of the particle moving along the surface of the system. Therefore, is natural to assume that the effective mass is related to the surface curvature. The surface curvature information is encoded in the second fundamental form $b_{ij}$ which satisfies \cite{spivak}
\begin{equation}
    \frac{\partial^2 \vec{r}}{\partial q^1 \partial q^j}=b_{ij}\hat{\mathbf{n}} + \Gamma^{k}_{ij}\frac{\partial\vec{r}}{\partial q^k},
\end{equation}
where $\hat{\mathbf{n}}$ in the unit normal vector to the surface and $\Gamma^{k}_{ij}$ are the Christoffel symbols.
Since the catenoid geometry is isotropic, we adopt a position-dependent mass in the form,
\begin{equation}
\label{generalpdm}
    m^{*}(x)=m^{*}f(H,K),
\end{equation}
being $m^*$ the effective mass of the electron in a flat graphene sheet and $f(H,K)$ is a particular function of the scalar invariants of the second fundamental form, the mean curvature $H=Tr(b_{ij})/2$ and gaussian curvature $K=det(b_{ij})$. The presence of the both curvatures $H$ and $K$ not only leads to the position dependence of the effective mass $m^*$ but also guarantees that intrinsic and extrinsic deformations of the surface modifies the electron effective mass. It is worthwhile to mention that the position dependent mass considered in Ref.\cite{moraes} satisfies Eq.\eqref{generalpdm} since the nanotube has $H=d^{-1}$ and $K=0$, and so $m^{*}(x)=m^{*}H^{\alpha}$.

In this work we consider an isotropic ansatz of form $m^{*}(x)=M(u)$ in the form 
\begin{equation}\label{m1}
    M(u)=m^{*}[1+\lambda(H^2-K)^\alpha],
\end{equation}
being $\lambda$ a parameter given by the system, called coupling factor. The function $f(H,K)=1+\lambda(H^2-K)^\alpha$ is dimensionless and it is build from the invariant term $(H^2-K)$, known in the elasticity theory as the Wilmore energy \cite{wilmore, Helfrich}. For $\alpha=1$, the coupling factor $\lambda$ has dimension of $length^2$ whereas the curvature term $(H^2 -K)$ has dimension of $length^{-2}$.
Note that for the case of the catenoid (\ref{m1}) gives us
\begin{equation}\label{mass}
    M(u)=m^{*}r(u),
\end{equation}
where the function $r(u)$ is defined by
\begin{equation}
r(u)=  1+ \frac{\lambda R^2}{(u^2+R^2)^2}.
\end{equation}
Note that $M(u)\rightarrow m^{*}$ when $u\rightarrow\pm\infty$, in agreement with the fact that the catenoid is asymptotically flat when $u\rightarrow\infty$. When $u\rightarrow 0$ then $M(u)\rightarrow m^{*}(1+\lambda/R^2)$. Thus, $\sqrt{|\lambda|}$ measures a length of influence of the position-dependent mass effects.

The coupling constant $\lambda$ can assume positive or negative values. For $\lambda>0$, the effective mass is augmented near the bridge throat and assumes the flat value $m^*$ asymptotically. For $\lambda<0$, the curvature reduces the effective electron mass near the bridge. In order to prevent a vanishing effective mass, the coupling constant $\lambda$ should satisfy $\sqrt{-\lambda}\neq R$.
For $\sqrt{-\lambda}< R$, the effective mass is everywhere positive. On the other hand, for $\sqrt{-\lambda}> R$ the effective mass is negative around the bridge throat interval $-R<u<R$ and positive for $|u|>R$. The effects of the coupling constant $\lambda$ on the bound states will be explore in details in the next section.

Using the curvature-dependent mass ansatz (\ref{mass}), the stationary Schr\"{o}dinger equation along the meridian (\ref{meridianschrodingerequation}) reads
\begin{eqnarray}
\label{freenonhermitionequation}
\nonumber&-&\frac{h^2}{2m^*}\left[\partial_u^2\Phi+\frac{u}{R^2 + u^2}\partial_u\Phi- \frac{\nu^2}{R^2 + u^2}\Phi\right]\\
\nonumber&&-\frac{\hbar^2}{6}\left[\partial_u^2\left(\frac{1}{m^*}\right)+\frac{u}{R^2+u^2}\partial_u\left(\frac{1}{m^*}\right)\right]\Phi\\
&&-\frac{\hbar^2}{2}\partial_u\left(\frac{1}{m^*}\right)\partial_u\Phi-\frac{\hbar^2}{2m^*}\frac{R^2}{(R^2+u^2)^2}\Phi=E\Phi
\end{eqnarray}
Note that the Eq.\ref{freenonhermitionequation} exhibits parity and time-reversal invariance, as a result of the catenoid geometric symmetries, as we can see from the acting of parity $\mathcal{P}\hat{u}(z)\mathcal{P}=\hat{u}(-z)=-\hat{u}(z)$ and time reversal $\mathcal{T}\hat{u}(z)\mathcal{T}=\hat{u}(z)$ operators upon $u(z)$. Nonetheless, the first order derivative terms renders the Hamiltonian non-Hermitian with respect to the meridian momentum $\hat{P}_u:= -i\hbar \partial_u$. The non-Hermiticity of the free electron Hamiltonian is not a problem, since the space-time reflection symmetry is preserved, the spectrum of the eigenvalues of the Hamiltonian is completely real \cite{Bender, Bender2}. Besides, there is an Hermitean equivalent Hamiltonian that can be achieved by a simple changing of variables. Considering the change on the wave function 
\begin{equation}
    \Phi(u)=\frac{1}{2} e^{\log  \left(R^4+\lambda  R^2+2 R^2 u^2+u^4\right)-\frac{5}{2} \left(\log  \left(R^2+u^2\right)\right)}y(u),
\end{equation}
leads to an one dimensional Hermitian Schr\"{o}dinger-like equation
\begin{equation}\label{modifiedschrodingerequation}
    -\frac{\hbar^2}{2M(u)}\frac{d^2y}{du^2} + V_{eff}(u)y=Ey,
\end{equation}
whose effective potential is given by
\begin{equation}
\label{freeeffectivepotential}
V_{eff}(u)=V_c(u)+V_{G}(u)+V_{\lambda}(u),
\end{equation}
where $V_c(u)$ is the centrifugal potential, $V_{G}(u)$ is the geometric potential and $V_{\lambda}(u)$ is the coupling potential, that are given by:
\begin{equation}
    V_c(u)=\frac{\hbar^2}{2M(u)}\left[\frac{\nu^2}{(R^2+u^2)}\right]
\end{equation}
\begin{equation}
    V_{G}(u)=-\frac{\hbar^2}{2M(u)}\left[\frac{(2R^2+u^{2})}{4(R^2+u^2)^2}\right]
\end{equation}
\begin{equation}
    V_{\lambda}(u)=\frac{\hbar^2}{2M(u)}\left\{\frac{2\lambda R^2[\lambda R^2(R^2-2u^2)+(R^2-4u^2)(R^2+u^2)^2]}{3[\lambda R^2(R^2+u^2)+(R^2+u^2)^3]^2}\right\}.
\end{equation}
Although the equation (\ref{modifiedschrodingerequation}) resembles a Schr\"{o}dinger equation, the operator on the left side $\hat{\mathcal{H}}y(u)=r(u)Ey(u)$ is not properly the Hamiltonian, since the "eigenvalue" depends on $u$. By performing another wave function redefinition as $y(u)=\frac{s_0}{\sqrt{r(u)}}\chi(u)$, being $s_0$ and arbitrary constant, we obtain the more familiar PDM Schr\"{o}dinger equation \cite{pdm1,pdm2,pdm3,pdm4}
\begin{eqnarray}\label{Sturmform}
-\frac{\hbar^2}{2}\frac{d}{du}\left[\frac{1}{M(u)}\frac{d\chi}{du}\right]+\widetilde{V}_{eff}(u)\chi=E\chi,
\end{eqnarray}
and the effective potential is rewritten as
\begin{eqnarray}\label{Vbar}
\widetilde{V}_{eff}(u)=V_{c}(u)+V_{G}(u)+\widetilde{V}_{\lambda}(u),
\end{eqnarray}
since, due to the redefinition of the wave function $y(u)$, the coupling potential is rewritten as
\begin{eqnarray}\label{vbar1}
\widetilde{V}_{\lambda}(u)=V_{\lambda}(u)+\frac{\hbar^2}{2M(u)}\left(-\frac{1}{2}\frac{r''(u)}{r(u)}+\frac{3}{4}\left(\frac{r'(u)}{r(u)}\right)^2\right).
\end{eqnarray}

Thus, the electron's dynamics on a catenoid with Hamiltonian  (\ref{freenonhermitionequation}) is Hermitian equivalent to an electron under the action of the effective potential in (\ref{Vbar}). Moreover, the  effects of curvature, PDM and angular momentum are encoded in the effective potential. The equivalence between a $\mathcal{P}$ $\mathcal{T}$ symmetric non-hermitean Hamiltonian and a Hermitean Hamiltonian has attracted much attention in last years \cite{Jones, Andrianov1, Andrianov2}. Moreover, since (\ref{Sturmform}) satisfies the continuity equation  $\frac{\partial |\chi|^2}{\partial t} + \frac{\partial J_u}{\partial u}=0$, the square of the wave function $|\chi|^2$ represents a true probability density.

Asymptocially, i.e. for $u\rightarrow \pm\infty$,
the effective potential in Eq. $\ref{Vbar}$ vanishes and the PDM function reaches its asymptotic value $M(U)\rightarrow m^{*}$, so that the respective asymptotic solution of Eq.($\ref{Sturmform}$) are of form
\begin{equation}
\chi(u)\approx A\cos(k u + \varphi),
\end{equation}
where $k^2=\frac{2 m^{*} \varepsilon}{\hbar^2}$. Therefore, far from the bridge throat where the curvature effects vanish the electron behaves as free states.
In the next section, we analyze the effects of the curvature and PDM on bound states near the bridge throat.

\subsection{Qualitative analysis}
Before obtain the bound states and their respective spectra, let us discuss some qualitative features of the effective potential.

\begin{figure}[h!]
\begin{center}
\includegraphics[scale=0.6]{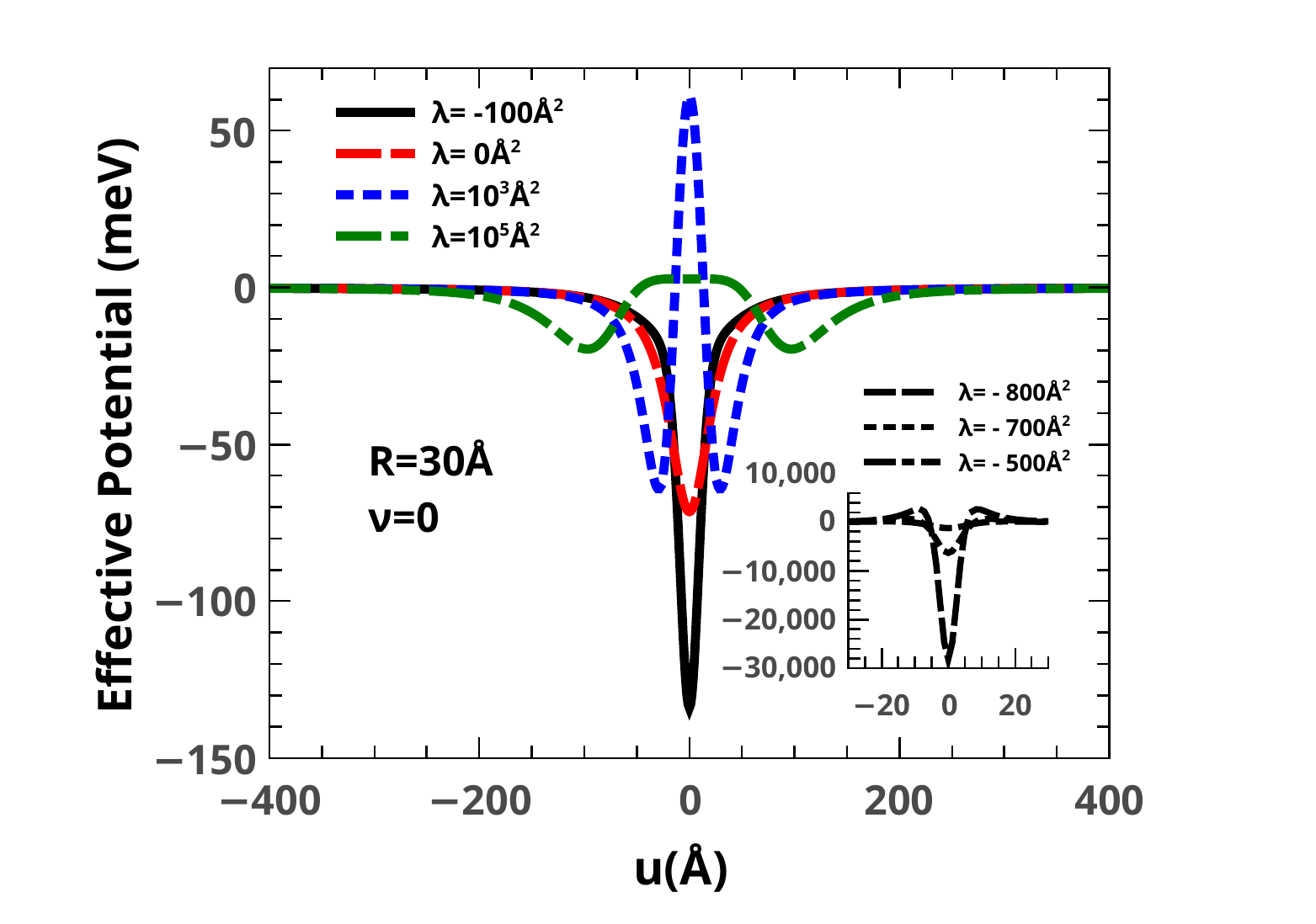}  
\caption{The effective potential for $R=30\AA$ and $\nu=0$ for some values of $\lambda$, obtained by the eq.\ref{Vbar}.}
\label{FIG2}
\end{center}
\end{figure}

In the fig.\ref{FIG2}, we show the effective potential given by the eq.\ref{vbar1}, for $R=30$ \AA, and $\nu=0$, for some values of $\lambda$. The eq.\ref{mass} shows that the mass depends on coupling factor, $\lambda$, and that in addition, $\lambda$ can assume negative values, but not any negative value, there is a minimum limit value, which we call a critical coupling factor, $\lambda_{c}$, from which the mass becomes negative, this critical value is given by $\lambda_{c}=-R^{2}$. For $R=30$ \AA, the critical coupling factor é $\lambda_{c}=-900$ \AA$^{2}$. In the insert of the figure, we observe a deep well for $\lambda=-800$ \AA$^{2}$, and it becomes deeper when $\lambda$ assumes values close to the critical value, $\lambda=\lambda_{c}$, 
in this case the coupling effect suppresses the geometric effect. 
The effective potential decreases its depth when the $\lambda$, having negative values, is increased until it reaches zero, as shown in the figure. A single well is observed for $\lambda=0$, this is the case when the effective mass is the same over the entire catenoid, $M(u)=m^{
*}=0.03m_{0}$. \cite{euclides} 
When $\lambda$ assumes positive values, the effective potential is drastically altered, generating two wells symmetrical in relation to the origin of the catenoid, in addition the double well becomes softer when lambda takes on higher values, this behavior is shown in the results taken for
for $\lambda=10^{3}$ and $10^{5} $ \AA$^{2}$.   
Under these conditions the coupling is so intense that the geometric effect is suppressed.  Therefore, the geometric effect is suppressed by the coupling effect in two limits, when $\lambda$ tends to $\lambda_{c}$ and when $\lambda$ tends to infinity.

\begin{figure}[h!]
\begin{center}
\includegraphics[scale=0.6]{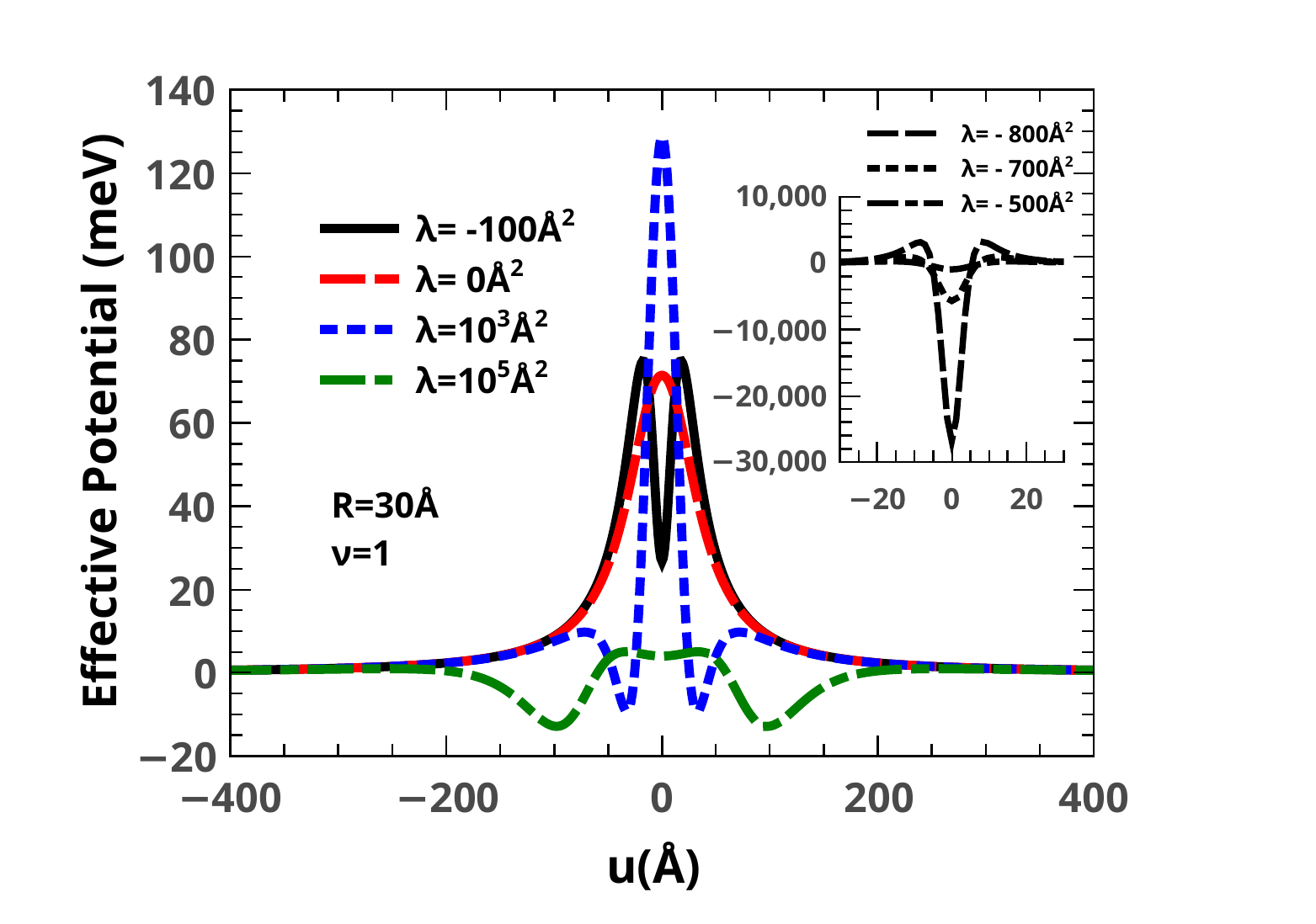}  
\caption{The effective potential for $R=30\AA$ and $\nu=1$ for some values of $\lambda$, obtained by the eq.\ref{Vbar}.}
\label{FIG3}
\end{center}
\end{figure}

The effective potential for $R=30 \AA$ and $\nu=1$ is shown in the fig.\ref{FIG3}. In this figure, in addition to the competition between the geometric and coupling potentials, discussed in fig.\ref{FIG2}, there is the centrifugal potential that significantly modifies the effective potential.
 When comparing the figures inserted in figs.\ref{FIG2} and \ref{FIG3}, we observe that the centrifugal effect is suppressed for $\lambda=-800$\AA$^{2}$, because the effective potentials have practically no changes. 
However, when $\lambda$, having negatives values,  
approaches zero, the centrifugal term becomes more relevant than the effect of coupling and a barrier, instead of a well, appears at the origin of the catenoid, as can be seen in fig.\ref{FIG3} for $\lambda=-100$ \AA$^{2}$. For $\lambda=0$ the effective mass is constant, $M(u)=m^{*}$, and the effective potential takes the form of a barrier \cite{euclides}. We also observed that when $\lambda$ assumes large positive values, the coupling effect changes the effective potential to form a double well symmetrical in relation to the origin of the catenoid, suppressing the geometric and centrifugal effects. This behavior is shown in the fig.\ref{FIG3} for $\lambda=10^{5}$ \AA$^{2}$, the same is observed in fig.\ref{FIG2} for the same value of $\lambda$. Here, it is worth noting that both the geometric effect and the centrifugal effect are suppressed by the two-limit coupling effect, when lambda tends to $\lambda_{c}$ and when lambda tends to infinity, as discussed in the fig.\ref{FIG2}.

\begin{figure}[h!]
\begin{center}
\includegraphics[scale=0.6]{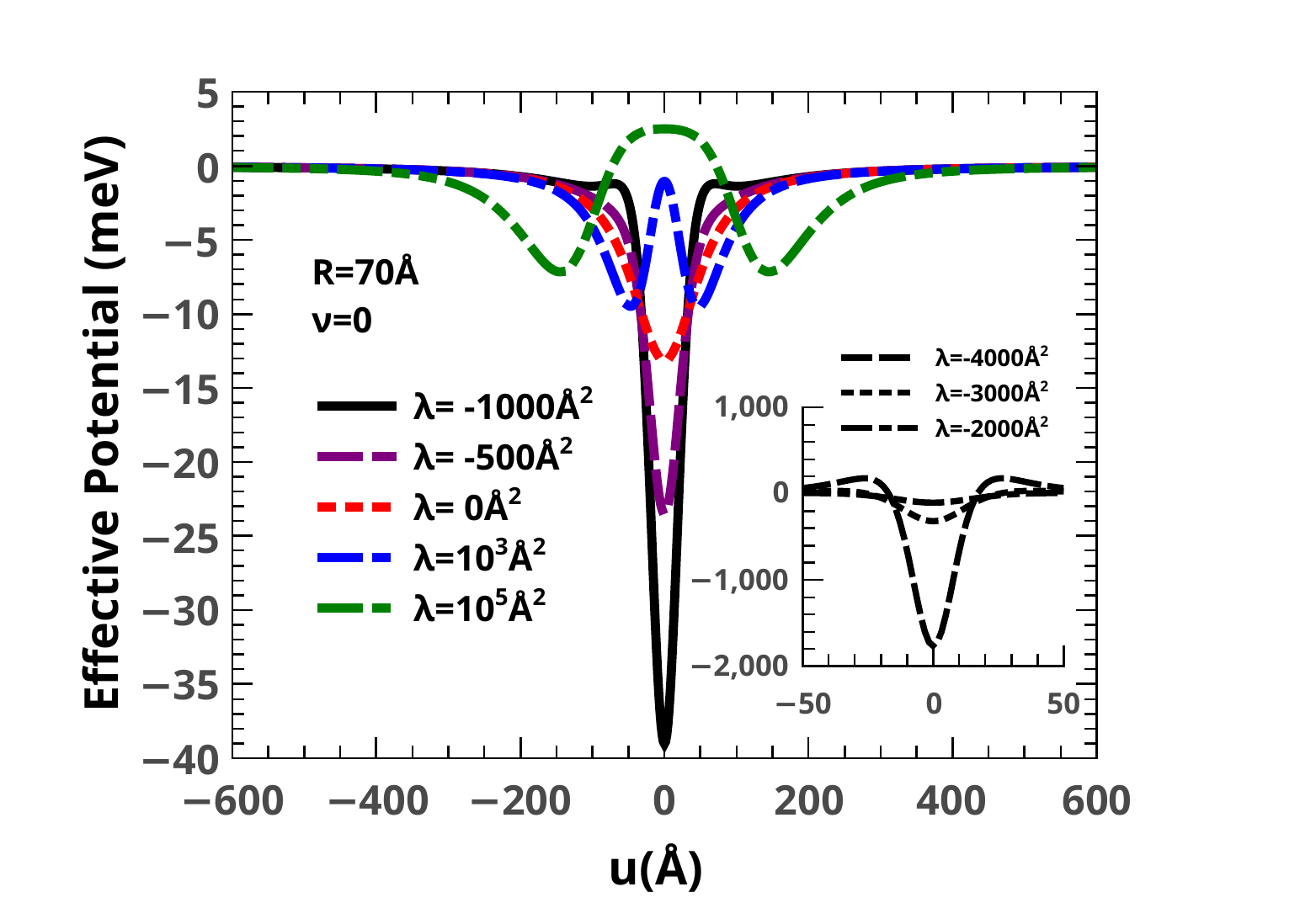}  
\caption{The effective potential for $R=70\AA$ and $\nu=0$ for some values of $\lambda$, obtained by the eq.\ref{Vbar}.}
\label{FIG4}
\end{center}
\end{figure}

The effective potential for $\nu=0$ and $R=70 \AA$ is shown in the fig.\ref{FIG4}, for some values of $\lambda$. The increase in the radius of the catenoid throat increases the domain of the lambda values, considering that the minimum value of the critical lambda decreases to $-4900\AA^{2}$. 
The behavior of the effective potential shown in fig.\ref{FIG4} is similar to that of fig.\ref{FIG2}, that is, the potential has a deep well in the origin of the catenoid for $\lambda$ values close to the critical value. The depth of the well decreases when $\lambda$, having negative values, has its value increased to zero and finally when $\lambda$ increases, having its positive values, a double well symmetrical in relation to the origin of the catenoid appears, and the increase in $\lambda$ decreases the depth of the double well. In both cases the centrifugal potential is not present ($\nu=0$), so the geometric potential competes with the coupling potential. Although qualitatively the results presented in fig.\ref{FIG4} and \ref{FIG2} are similar, the radius of the catenoid throat is different. In this condition, increasing the radius value, $R$, decreases the geometric confinement on the effective potential, as studied in our recent work, see ref. \cite{euclides}, as the critical lambda, $\lambda_{c}$, depends on the radius, the increase in the radius also causes the coupling potential to decrease its influence on the effective potential, but more smoothly, this is more evident when we compare the effective potentials with $\lambda$ values close to their respective critical lambdas, see the effective potential for $\lambda= -800 \AA^{2}$ in the figure inserted in fig.\ref{FIG2}, and for $\lambda = -4000\AA^{2}$ in the figure inserted in fig.\ref{FIG4}.

\begin{figure}[h!]
\begin{center}
\includegraphics[scale=0.6]{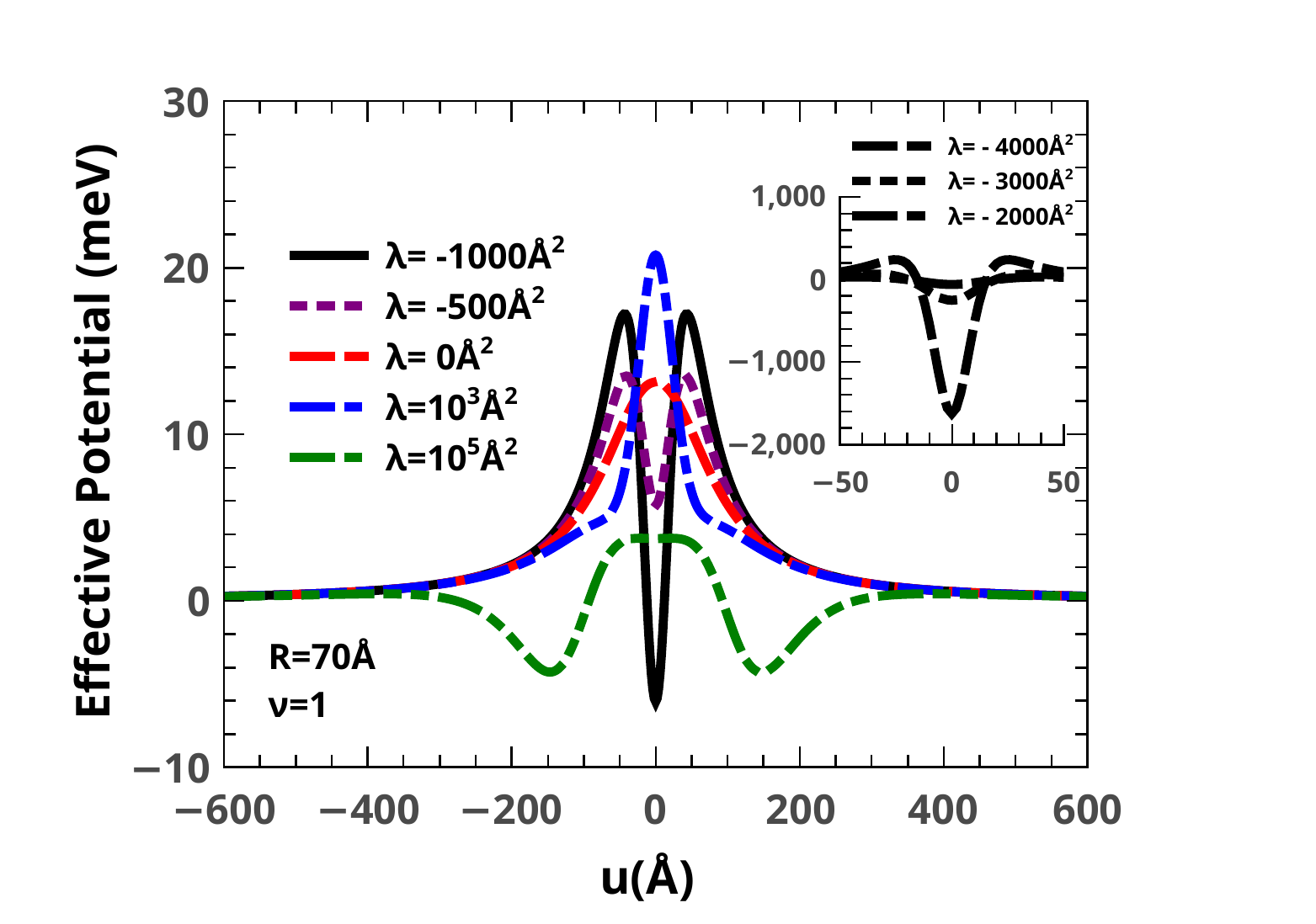}  
\caption{The effective potential for $R=70\AA$ and $\nu=1$ for some values of $\lambda$, obtained by the eq.\ref{Vbar}.}
\label{FIG5}
\end{center}
\end{figure}
The effective potential for $\nu=1$ and $R=70\AA$ is shown in the fig.\ref{FIG5}. This effective potential presents a qualitative behavior similar to the effective potential shown in fig.\ref{FIG3}. In both cases, the centrifugal potential is present ($ \nu = 1 $), and the results show that the increase in the radius, $R$, decreased both the geometric effect and the coupling effect on the effective potential, see the effective potential for $\lambda= -800 \AA^{2}$ in the figure inserted in fig.\ref{FIG3}, and for $\lambda = -4000\AA^{2}$ in the figure inserted in fig.\ref{FIG5}.



\section{Bound states}
\label{section3}

In the previous section, we discussed the effective potential generated by the surface of a catenoid of radius $R$, made of graphene, in which the effective mass of the electron depends on a coupling factor $\lambda$. In this section we will look at the electronic states accessible in these potentials.

For this, we solve numerically the eq.\ref{Sturmform}, using the finite difference method \cite{Ramos11}, for the effective potential given by the eq.\ref{Vbar}, for some values of $R$ and $\lambda$. In the calculations, we use $m^{*} = 0.03 m_{0}$, which is the effective mass the electron on a single layer graphene sheet and $m_{0}$ is the resting mass of the electron.  
\begin{figure*}[ht!]
\begin{center}
\includegraphics[scale=0.8]{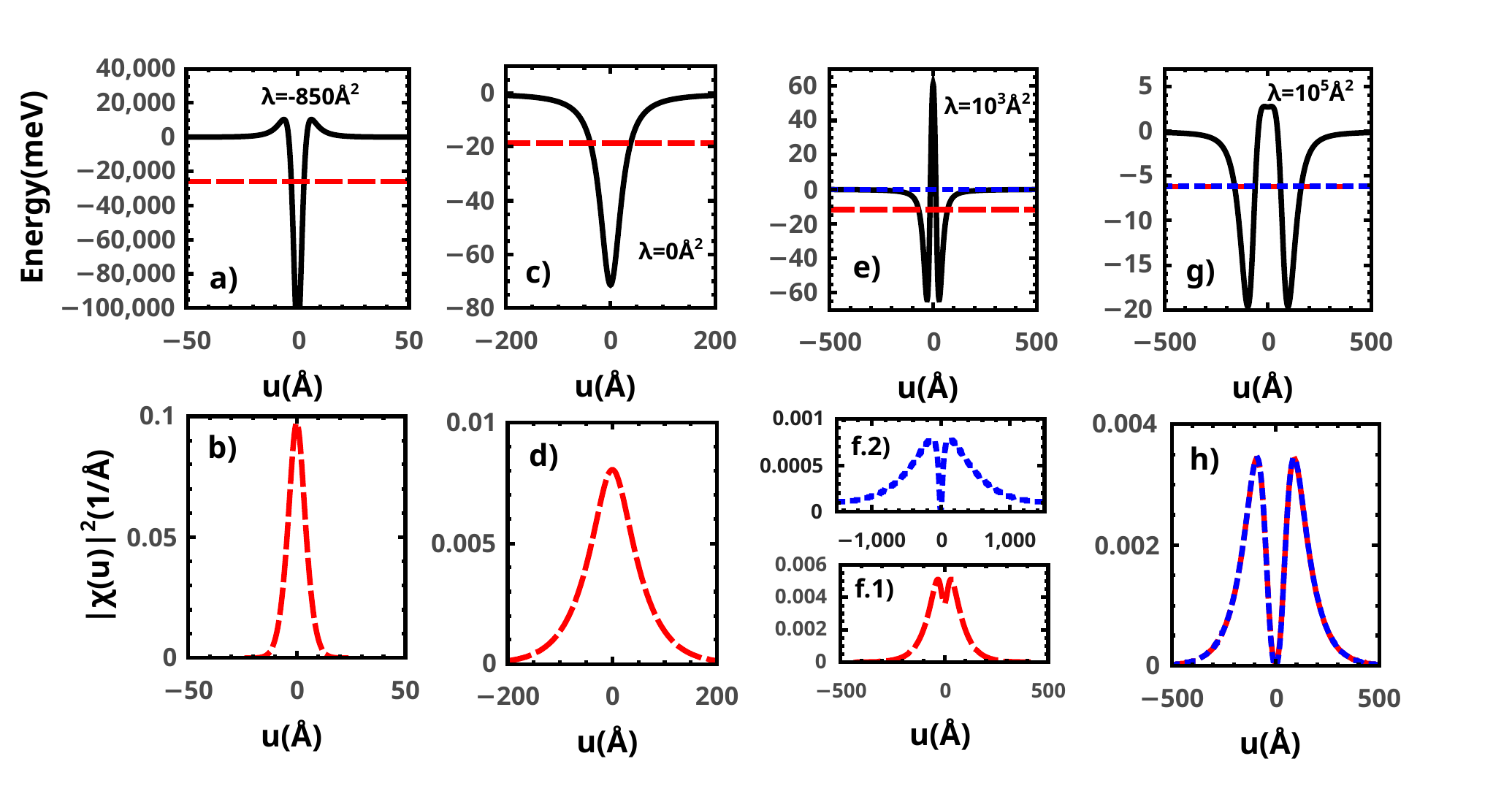}   
\caption{The bound states and their probability densities for a catenoid with radius $R = 30 \AA$ and $\nu = 0$. The solid black line represents the effective potential for: a) $\lambda = -850 \AA^{2}$, c) $\lambda = 0 \AA^{2}$, e) $\lambda = 10^{3} \AA^{2}$ and g) $\lambda = 10^{5} \AA^{2} $. The red dashed and blue dotted lines correspond to the first and second bound states and their probability densities, respectively.}
\label{FIG6}
\end{center}
\end{figure*}

The fig.\ref{FIG6} shows the bound states for $R=30 \AA$ and $\nu=0$ for four values of $\lambda$. We observe that the electronic state more confined is given by the effective potential for $\lambda = -850$ \AA$^{2}$ shown in the fig.\ref{FIG6} a), the energy of this state is -26,020.21 meV and the probability density function shows that the electron has high probability of being found close to the origin of the catenoid, see the fig.\ref{FIG6} b). Making use of the angular symmetry of the catenoid and taking the width of the half height of the probability density function, which is $\Delta u=8.6$ \AA, we can visualize a probability cloud around the origin of the catenoid in the form of a ring, which we can call a probability ring. From what we see, this state is very localized, this is because the coupling factor chosen is very close to $\lambda_{c}$, in this situation the electron has a very small effective mass.

The probability ring is wider, $\Delta u=100.8$ \AA, for $\lambda = 0$ \AA$^{2}$, so in this condition the electron is less confined, as shown by the value of the energy level of the bound state which is -18.63 meV, as shown in fig.\ref{FIG6} c) and d).

Two confined energy states appear for $\lambda = 10^{3}$ \AA$^{2}$, the first state is -11.70 meV and the second is -0.24 meV, see fig.\ref{FIG6} e). These states are not Gaussian functions, but states, called hybrids, that arise from the mixture of the states of the two wells symmetrical in relation to the origin of the catenoid \cite{Studart}. The first state (dashed red line) presents a probability density function in the form of two Gaussians, practically overlapping, with their maximum separated from $\Delta u = 64.8$ \AA, one located at $u = -32.4$ \AA, and the other located at $u = 32.4$ \AA, see the fig.\ref{FIG6} f.1). The cloud of probability associated with these states are in the form of two rings very close symmetrical in relation to the origin of the catenoid.

The second state (dotted blue line) presents also a probability density function in the form of two Gaussians, practically overlapping, but with their maximum separated from $\Delta u = 211.2$ \AA, one located at $u = -105.6$ \AA, and the other located at $u = 105.6$ \AA, see the fig.\ref{FIG6} f.2). The cloud of probability associated with these states are also in the form of two rings very close symmetrical in relation to the origin of the catenoid, but wider which represents that this state is less confined in relation to the first.

Now, for $ \lambda = 10^{5}$ \AA$^{2} $ these two hybrid states are practically degenerate, as they have almost the same energy value, $ -6.19 $ and $ -6.17 $ meV, see fig.\ref{FIG6} g), the probability density functions of the first hybrid state (red dotted line) and the second hybrid state (blue dotted line) are shown in fig.\ref{FIG6} h). The probability density function of each of these states takes the form of two Gaussians, one of the maximums being in $ u = -90 $ \AA, and the other is in $ u = 90 $ \AA.  Again, taking into account the angular symmetry of the catenoid and the probability density function of the first state (red dashed line), fig.\ref{FIG6} h), we can visualize a probability cloud in the form of two rings located symmetrically in relation to the origin of the catenoid, with the width of each of these rings being $\Delta u = 124.9$ \AA. This probability density function expresses the fact that
that an electron initially located in one of the wells can pass from one side to the other of the catenoid, as it presents a resonant tunneling. The same discussion holds for the second state as it has the same probability density function (blue dotted line) \cite{Studart}.

It is worth mentioning that the effective potential for $ \lambda = 10^{6} $ \AA$^{2}$ the energy of the bound states are -2.26 and -2.26 meV. This reflects what was said in the previous section, the effective potential becomes more shallow as the coupling factor increases, $ \lambda $, consequently the energy of the connected states gets smaller and smaller to the point of ceasing to exist. In these conditions the effective mass of the electron is so large that the effective potential can no longer confine it, so an electron coming from infinity, on the surface of the catenoid, passes through the throat of the catenoid without realizing its existence.

\begin{figure*}[ht!]
\begin{center}
\includegraphics[scale=0.8]{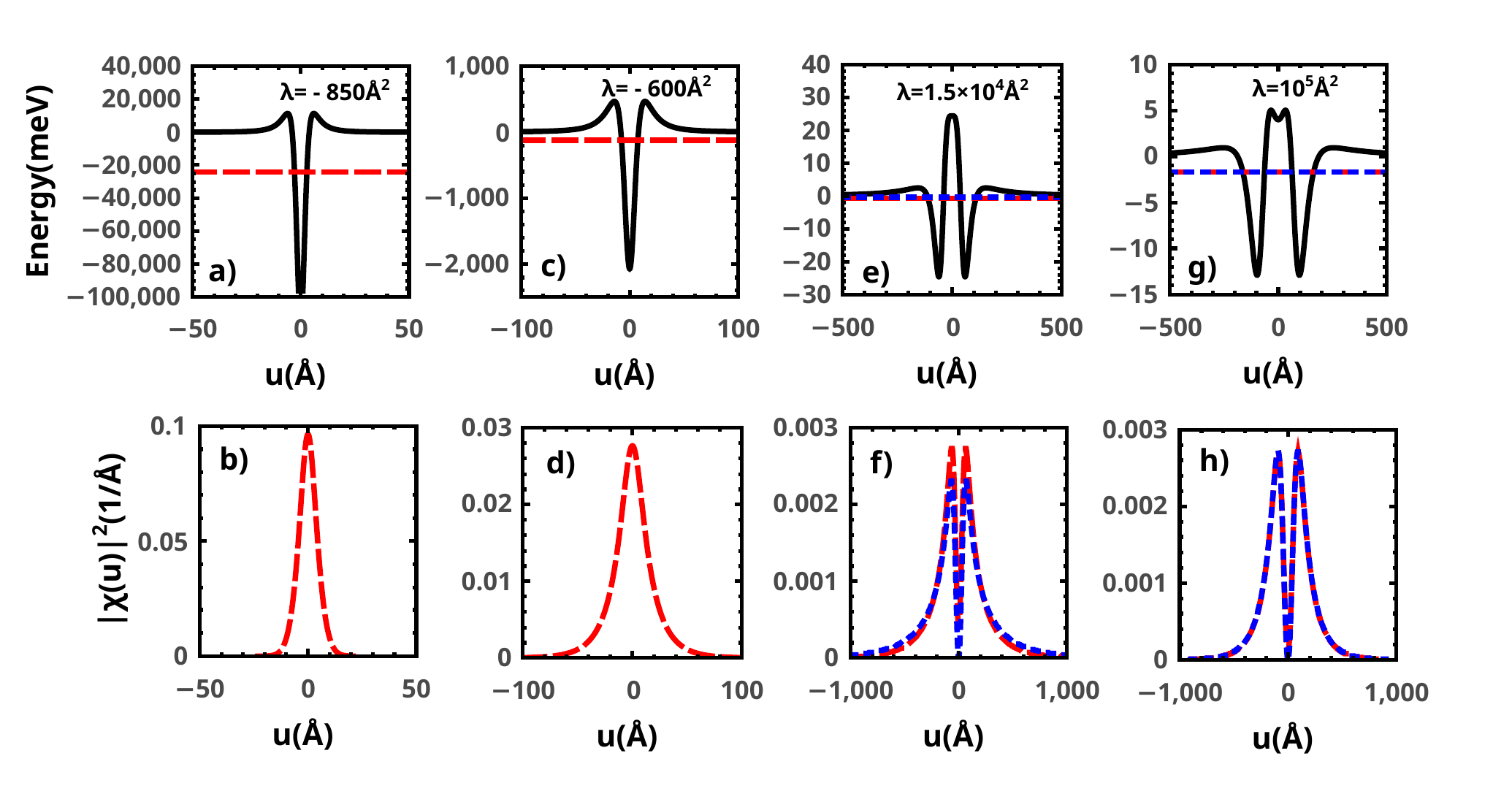}   
\caption{The bound states and their probability densities for a catenoid with radius $R = 30 \AA$ and $\nu = 1$. The solid black line represents the effective potential for: a) $\lambda = -850 \AA^{2}$, c) $\lambda = -600 \AA^{2}$, e) $\lambda = 1.5 \times 10^{4} \AA^{2}$ and g) $\lambda = 10^{5} \AA^{2} $. The dashed red and blue dotted lines correspond to the first and second bound states and their probability densities, respectively.}
\label{FIG7}
\end{center}
\end{figure*}

The fig.\ref{FIG7} shows the bound states for $R=30 \AA$ for four values of $\lambda$, however, we take in account the centrifugal potential, $\nu = 1 $. We observe that the states become less localized when compared with the results, see the fig.\ref{FIG6}, where the centrifugal potential is absent.
When looking at figs.\Ref{FIG6} a) and \ref{FIG7} a), we do not observe the influence of the centrifugal potential on the effective potential, because apparently the potentials are very similar, and even the probability density functions have, approximately, the same width ($\Delta u = 9.2 $ \AA), however, the energy value of the bound state is changed to -24,206.83 meV.

The fig.\ref{FIG7} c) shows the energy of the bound state, for $\lambda = -600 $ \AA$^{2}$, which is -123.70 meV, and the probability ring, located at the origin, has a width of $ \Delta u = 28$ \AA, as shown the fig.\ref{FIG7} d). We observed that for values of $\lambda$ closer to $\lambda_{c}$, the electronic state is more confined.

No bound state is observed for $\lambda=0$ \AA$^{2}$, because for $\nu=1$, the effective potential is a barrier and not a well. 
This barrier is shaped like a Gaussian, whose height is 71.43 meV and its width is 35.71 meV \cite{euclides}.

Due to the presence of the orbital angular momentum, $\nu = 1$, the linked states were only obtained for values of $\lambda$ greater than $10^{4}$ \AA$^{2}$, then we calculate the effective potential for $\lambda = 1.5\times 10^{4} $, for this configuration, two hybrid states are observed as shown in the fig. \ref{FIG7} e) and f), the energy of the first (dashed red line) and second (dotted blue line) states are are -0.72 meV and -0.33 meV, respectively. The probability cloud, of the first state (dashed red line), is in the form of two rings, one located at $u= -63.6 $ \AA\, and the other located at $ u=63.6 $ \AA, each ring having a width of $\Delta u=134.8$ \AA. While the second (dotted blue line), the probability rings are located in $u=-66.0$ \AA\, and $u=66.0$ \AA, and the width of each one is $\Delta u =147.6$ \AA.

Two hybrid states also appear for $ \lambda = 10^{5} $, fig.\ref {FIG7} g) and h), the energy of the first (dashed red line) and second (dotted blue line) states are -1.69 meV and -1.66 meV, respectively. Their probability rings are practically coincident and they are located at $ u = -92.4 $ \AA\, and $ u = 92.4 $ \AA, with the width of the rings being $\Delta u = 150 $ \AA. These states are less confined compared to the states shown, in figs. \ref{FIG6} g) and h), where the centrifugal potential is absent, $\nu=0$.

\begin{figure*}[ht!]
\begin{center}
\includegraphics[scale=0.8]{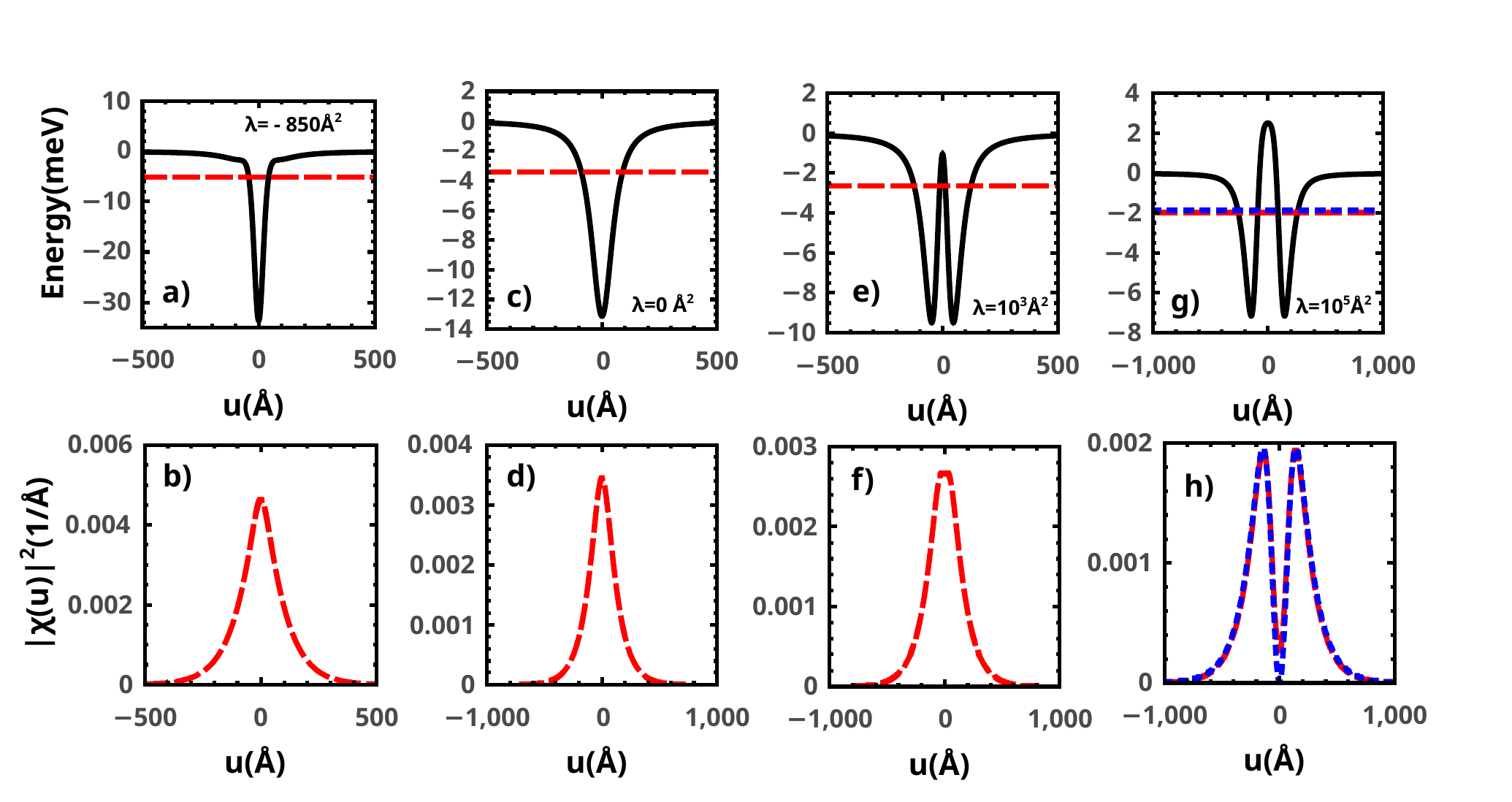}   
\caption{The bound states and their probability densities for a catenoid with radius $R = 70 \AA$ and $\nu = 0$. The solid black line represents the effective potential for: a) $\lambda = -850 \AA^{2}$, c) $\lambda = 0 \AA^{2}$, e) $\lambda = 10^{3} \AA^{2}$ and g) $\lambda = 10^{5} \AA^{2} $. The dashed red and blue dotted lines correspond to the first and second bound states and their probability densities, respectively.}
\label{FIG8}
\end{center}
\end{figure*}

The fig.\Ref{FIG8} shows the bound states and their probability density functions for $\nu=0$ and $R=70$ \AA, for four values of the $\lambda$. According to the results shown in fig.\ref{FIG8}, the increase in the radius of the catenoid decreases and even remove the bound state, as is the case for $\lambda=10^{3}$ \AA$^{2}$. So, increasing the radius of the catenoid decreases the effect of the geometric potential on the effective potential of the system. For example, for $\lambda=-850$ \AA$^{2}$, the energy of the bound state is -5.14 meV, and the width of the probability density function is $\Delta u=168.4 $ \AA. 

For $\lambda=0$ \AA$^{2}$ the energy of the bound state is -3.42 meV, and the width of the probability density function is $\Delta u=236.8$ \AA. Now for $\lambda=10^{3}$ \AA$^{2}$, one of the states seen in the fig.\Ref{FIG6} e) is removed, and the remaining one has energy of -2.63 meV, with the width of the bound state being $\Delta u=312.0$ \AA, see the fig.\Ref{FIG8} f).

\begin{figure*}[ht!]
\begin{center}
\includegraphics[scale=0.8]{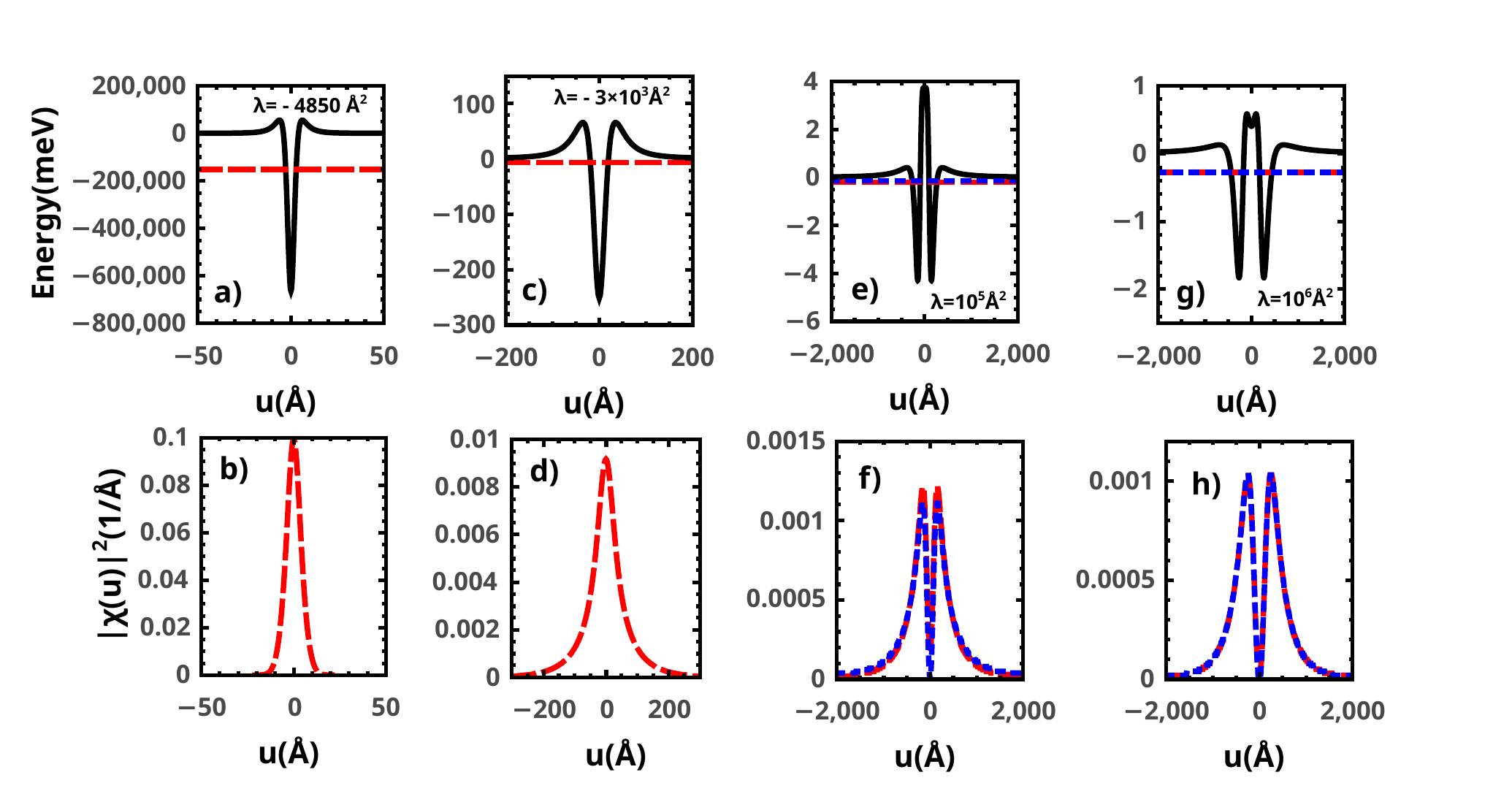}   
\caption{The bound states and their probability densities for a catenoid with radius $R = 70 \AA$ and $\nu = 1$. The solid black line represents the effective potential for: a) $\lambda = -4850 \AA^{2}$, c) $\lambda = -3000 \AA^{2}$, e) $\lambda = 10^{5} \AA^{2}$ and g) $\lambda = 10^{6} \AA^{2} $. The dashed red and blue dotted lines correspond to the first and second bound states and their probability densities, respectively.}
\label{FIG9}
\end{center}
\end{figure*}

The same hybrid states that appear for $R=30$ \AA\, and $\lambda=10^{5}$ \AA$^{2}$, for both $\nu=0$ and $\nu=1$, see the figs.\Ref{FIG6} and \ref{FIG7}, also appear for $R=70$ \AA, as shown in fig.\ref{FIG8} g) and h), however they are less localized. The energy of the first state (dashed red line) is -1.97 meV, and of the second (dotted blue line) is -1.87 meV. Here we notice a slight breakdown of degeneracy, although the probability density functions, which are shown in the fig.\Ref{FIG8} h), are overlapping. From the probability density functions we can visualize two probability rings, one located at $u=-140$ \AA and the other at $u=140$ \AA, being the width of each of these rings $\Delta u=219.2$ \AA. These hybrid states become even less localized to take into account the orbital angular momentum, $\nu=1$, for $R=70$ \AA, as shown in the fig.\Ref{FIG9} g) and h), in which the energy of the first state (dashed red line) is -0.19 meV and the second state (dotted blue line) is -0.13 meV. The probability cloud of the first state (dashed red line) is in the form of two rings, one located at $ u = -153.6 $ \AA\, and the other located at $ u = 153.6 $ \AA, each having a width of $ \Delta u = 299.2$ \AA. For the second state (dotted blue line), the cloud of probability is also shaped like two rings, one located at $ u = -158.4 $ \AA\, and the other located at $ u = 158.4 $ \AA, each having a width of $ \Delta u = 313.6 $ \AA.

No state linked to $\lambda=-850$ \AA$^{2}$ is found when in addition to increasing the radius of the catenoid to 70 \AA, we take into account the centrifugal potential, $\nu=1$.

The energy of the bound state for $\nu=1$, $R=70$ \AA, and $\lambda=-4850$\AA$^{2}$ is -152,812.01 meV, and the probability density function has a width equals to $\Delta u = 9.2 $ \AA. Both the energy of the bound state and its probability density function are shown in the figures \Ref{FIG9} a) and b). As we can see, this state is strongly linked because the value of $\lambda$ used is close to $\lambda_{c}=-4900$ \AA$^{2}$. The absence of the centrifugal potential ($\nu=0$) alters the energy of the bound state, to -154,638.10 meV, even for a coupling potential so close to the critical $\lambda_{c}$.

Increasing $\lambda$ to -3000 \AA$^{2}$ the energy of the bound state is altered to -6.49 meV, as well as, its probability density function, the width is $\Delta u=76$\AA, as shown the figs.\Ref{FIG9} c) and d).

Finally, for $\lambda=10^{6}$ \AA$^{2}$, the energy of the first state (dashed red line) is -0.274 meV and the second state (dotted blue line) is -0.273 meV. The probability cloud of the first state (dashed red line) and the second state (dotted blue line) is overlapping. The two probability clouds are shaped like two rings, one located at $ u = -248.0 $ \AA\, and the other located at $ u = 248.0 $ \AA, each having a width of $ \Delta u = $ 395.2 \AA. The question of these states becoming less and less localized when the coupling factor increases, has been discussed previously in the text.


\section{Final Remarks and perspectives}
\label{remarks}

We investigated the electronic states of a position-dependent mass (PDM) electron confined on the surface of a graphene catenoid bridge. In addition to the usual geometric potential, we considered a PDM mass as a function of the mean and the gaussian curvatures. 

The coupling between the curvature and the electron mass is controlled by a coupling constant $\lambda$, which can take any value in the interval $\lambda_{c}<\lambda<\infty$, where the critical value $\lambda_c$ is given by $\lambda_{c}=-R^{2}$ and $R$ is the radius of the caternoid throat.



The effective potential contains the usual geometric da Costa potential, the centrifugal term and the PDM corrections. Asymptotically, the effective potential vanishes and free states are allowed. Near the bridge throat, the effective potential is strongly dependent on the PDM parameter $\lambda$. For $\lambda_{c}<\lambda<0$, the potential exhibits an attractive volcano-like shape around the origin, regardless the value of the angular momentum.

For $0\leq \lambda<\infty$ the is rather dependent on the angular momentum. For $\nu=0$, $\lambda=0$ leads to a reflectionless potential \cite{euclides} and as $\lambda$ increases the potential takes the form of a double-well potential near the bridge throat. For $\nu=1$, the centrifugal term enhances the barrier near the origin.

The PDM parameter also modifies the number and behaviour of the bound states. For $\lambda_c <\lambda <0$, only one bound state located around the origin was found. As $\lambda$ increases keeping the volcano-shape potential the width of the bound state also increases. For $\lambda>0$, the PDM parameter breaks the bound state degeneracy leading to two states. Nonetheless, for greater values of $\lambda$, the double-well symmetry restore the degeneracy. In fact, there are two hybrid states, and these states are degenerate \cite{Studart}. The probability density function of these states indicates that resonant tunneling is possible. Our results also indicate that if the coupling factor continues to increase, that double potential well becomes very shallow and the bound states are suppressed.

As future developments we point out the analysis of the transport features of this graphene bridge, as well as the interaction of the confined electrons with electromagnetic fields.







\section*{Acknowledgments}
\hspace{0.5cm} J.E.G.Silva thanks the Conselho Nacional de Desenvolvimento Cient\'{\i}fico e Tecnol\'{o}gico (CNPq), grants n$\textsuperscript{\underline{\scriptsize o}}$ 312356/2017-0 for financial support.


\end{document}